\documentclass[11pt,a4paper]{article}
\usepackage{authblk}
\usepackage{anyfontsize}
\usepackage{tabularx}
\usepackage{multirow}

\usepackage[hyperref]{emnlp-ijcnlp-2019}
\usepackage{times}
\usepackage{latexsym}
\usepackage{graphicx}
\usepackage{amsmath}
\usepackage{array}
\usepackage{booktabs}
\usepackage{url}
\usepackage{amssymb}

\aclfinalcopy 

\title{Deep Multimodal Image-Text Embeddings\\
	for Automatic Cross-Media Retrieval\vspace{0.9em}}

\author[ 1]{\textbf{Hadi {Abdi Khojasteh}}}
\author[ 1, 2]{\textbf{Ebrahim Ansari}}
\author[ 1, 3]{\textbf{Parvin Razzaghi}}
\author[ 4]{\textbf{Akbar Karimi}}
\affil[1 ]{{\fontsize{10.5}{13.2}\selectfont Institute for Advanced Studies in Basic Sciences (IASBS), Zanjan, Iran}}
\affil[2 ]{{\fontsize{10.5}{13.2}\selectfont Faculty of Mathematics and Physics, Institute of Formal and Applied Linguistics, Charles University, Czechia}}
\affil[3 ]{{\fontsize{10.5}{13.2}\selectfont Institute for Research in Fundamental Sciences (IPM), Tehran, Iran}}
\affil[4 ]{{\fontsize{11.0}{13.2}\selectfont IMP Lab, Department of Engineering and Architecture, University of Parma, Parma, Italy}
	\authorcr\normalsize\texttt{{\fontsize{10.0}{12.6}\selectfont \{hkhojasteh,ansari,p.razzaghi\}@iasbs.ac.ir, akbar.karimi@unipr.it}}}

\date{}

\begin{document}
\maketitle
\hfill\vspace{0.5em}
\begin{abstract}
  This paper considers the task of matching images and sentences by learning a visual-textual embedding space for cross-modal retrieval. Finding such a space is a challenging task since the features and representations of text and image are not comparable. In this work, we introduce an end-to-end deep multimodal convolutional-recurrent network for learning both vision and language representations simultaneously to infer image-text similarity. The model learns which pairs are a match (positive) and which ones are a mismatch (negative) using a hinge-based triplet ranking. To learn about the joint representations, we leverage our newly extracted collection of tweets from Twitter. The main characteristic of our dataset is that the images and tweets are not standardized the same as the benchmarks. Furthermore, there can be a higher semantic correlation between the pictures and tweets contrary to benchmarks in which the descriptions are well-organized. Experimental results on MS-COCO benchmark dataset show that our model outperforms certain methods presented previously and has competitive performance compared to the state-of-the-art. The code and dataset have been made available publicly.
\end{abstract}

\section{Introduction}
The advent of social networks has brought about a plethora of opportunities for everyone to share information online in the forms of text, image, video and so forth. As a result, there is a vast amount of raw data on the Net which could be helpful in dealing with many challenges in natural language processing and image recognition. Matching pictures with their textual descriptions is one of these challenges in which the research interest has been growing \citep{wang2018cnn+,eisenschtat2017linking,faghri2017vse++,lee2018stacked}. 

\begin{figure}
	\vspace{2em}
	\centering
	\includegraphics[width=0.485\textwidth]{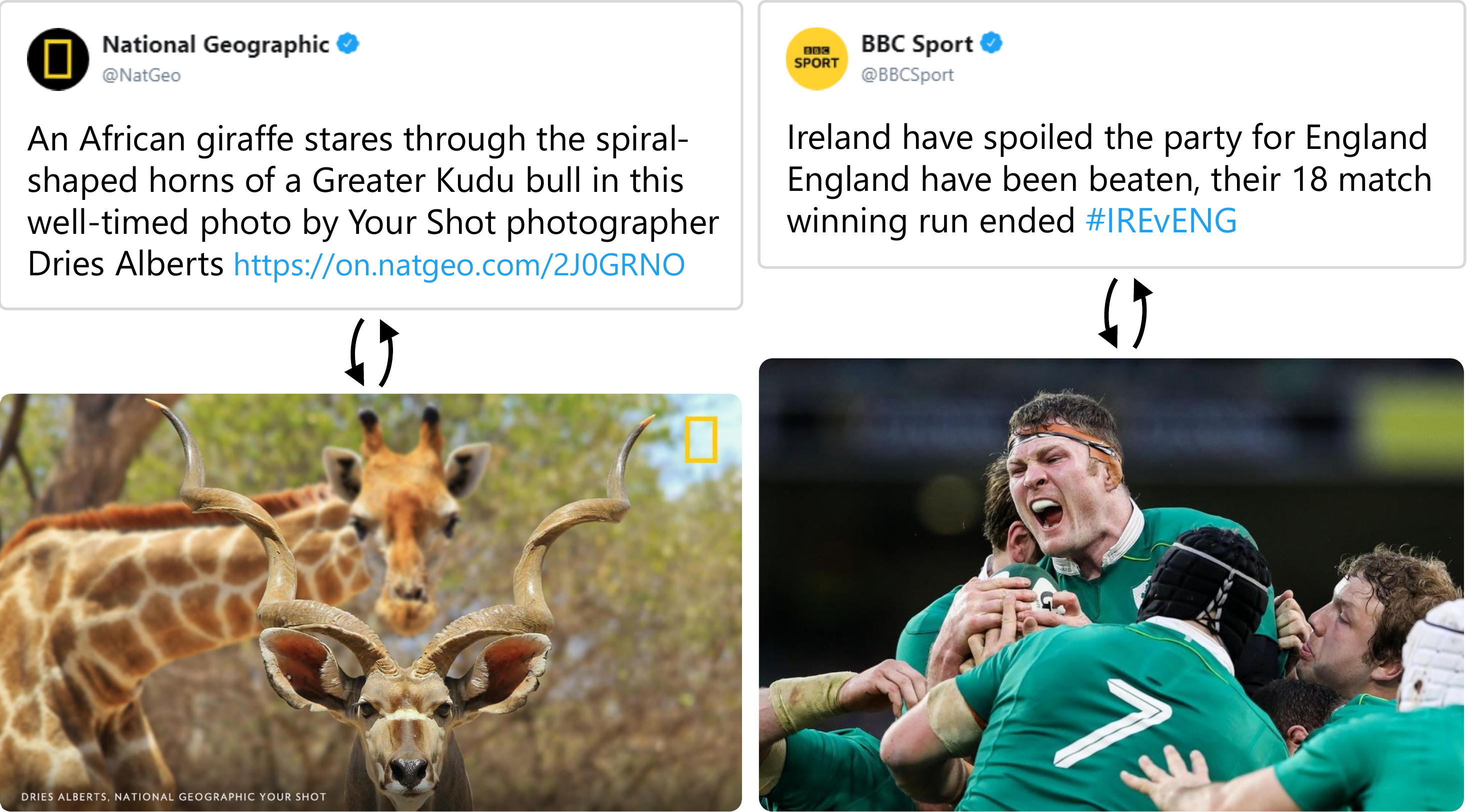}
	\caption{Motivation/Concept Figure: Given an image (caption), the goal in image-text matching is to automatically retrieve the closest textual description (image) for that. Tweets are examples of collected dataset.}
	\label{fig:fig1}
\end{figure}

The goal in image-text matching is, given an image, to automatically retrieve a natural language description of this image. In addition, given a caption (textual image description), we want to match it with the most related image found in our dataset as shown in Fig.~\ref{fig:fig1}. The process involves modeling the relationship between images and texts or captions used to describe them. This defines the semantics of a language by grounding it to the visual world.

Many studies have explored the task of cross-modal retrieval on the level of sentence and image regions \citep{wang2018cnn+,niu2017hierarchical,karpathy2015deep,liu2017learning}. \citet{karpathy2014deep} work on matching parts of an image objects with phrases by using dependency tree relations for sentence fragments and finding a common space for representing fragments. \citet{huang2017instance} propose a sm-LSTM where they utilize a multimodal context-modulated global attention scheme and LSTM to predict the salient instance pairs. Recently, many researchers \citep{huang2018learning,yan2015deep,zheng2017dual,donahue2015long,lev2016rnn,mao2014deep,gu2018look} introduced a neural network model for image caption retrieval consists of RNNs, CNNs, and additional multimodal layers. Practically, one of the reasons that these deep learning approaches have been on the rise is the availability of abundant information on the Web. The next section describes the proposed model.

\section{Model}
\begin{figure*}
	\centering
	\includegraphics[width=1.0\textwidth]{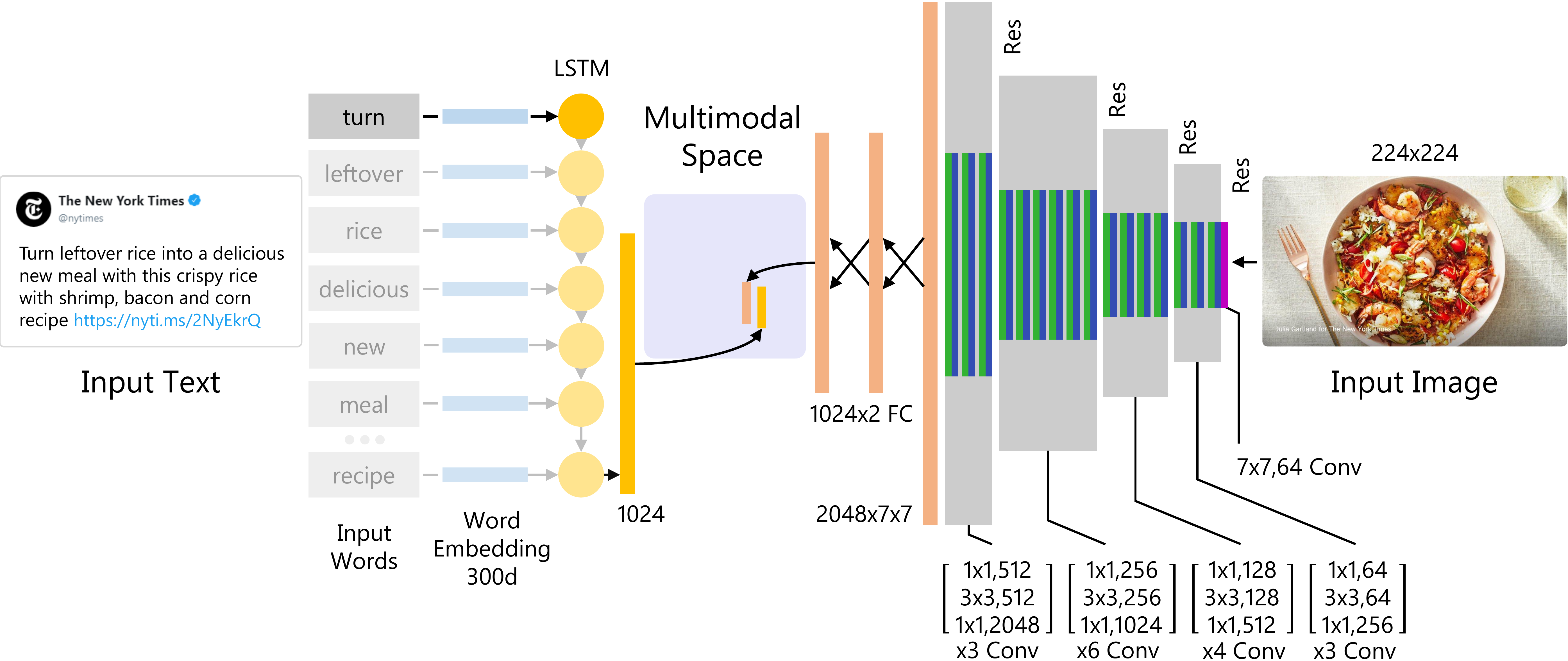}
	\caption{Proposed end-to-end multimodal neural network architecture for learning the image and text representations. Image features are extracted by a CNN with 16 residual blocks and text features are extracted by recurrent unit. Then the fully-connected layers join the two domains by feature transformation.}
	\label{fig:fig2}
\end{figure*}

In this work, we introduce an end-to-end multimodal neural network for learning image and text representations simultaneously. The architecture is illustrated in Fig.~\ref{fig:fig2}. It consists of two main subnets, a CNN for input image representation with an embedding and an LSTM to map the captions into the new space. The purpose of the model is to find a mapping from the text and image to a common space in order to represent them with similar embeddings. In this space, an image (text) will have a similar representation to its text (image) but a different one from other texts (images). Once the model is trained, by feeding an image (text) to the network, we find the most similar text (image).

\subsection{Image representation}
For our initial model, after removing the fully-connected layer from ResNet-50 \citep{xie2017aggregated} which has been pre-trained on ImageNet \citep{russakovsky2015imagenet}, we treat the remaining layers as an image feature extractor.
The inputs of the network are $224$$\times$$224$ images and the output is a $2048$$\times$$7$$\times$$7$ feature vector.
Therefore, a dense layer with the size of text domain is added to the end of the network. With the rest of the network, this layer which is now part of the model, is trained to produce image representations. If we call this vector $I$, which is a representation of the input image, then $f_{img}$ is a visual descriptor that is the result of forward pass in the network. The forward pass is denoted by $F_{img}(.)$, which is a non-linear function and is defined as $f_{img}=F_{img}(I)$. Conventionally, the image model is considered as one part of our network with its pre-trained weights to avoid computing a large number of learnable parameters which is a time-consuming process.
Then, we add two $1024$ fully-connected layers to transform $f_{img}(.)$ to an image feature vector ($v_{img}$) computed by $v_{img}$ = $W_{img}$$f_{img}(.)$ + $b_{img}$.

\subsection{Text representation}
Each input text ($T$) is first represented by an $n$$\times$$d$ matrix, with $n$ being its length and $d$ being the size of the dictionary. To build the dictionary, stop words and punctuation marks are removed and all the words are stemmed using porter stemmer. In addition, the removal of the special characters is carried out and the remaining words are all in lowercase format. Each word in the final dictionary is represented by a one-hot $d$ dimensional vector and every word can ﬁnd an index l in the dictionary. Therefore, for an input sentence $T$ with $m$ words, there is a $d$$\times$$m$ matrix as the following:

\begin{small}$T(i,j) =
\left\{
	\begin{array}{ll}
		1  & \mbox{} \hspace{15pt} j=l_i \\
		0 & \mbox{} \hspace{15pt} otherwise
	\end{array}
\right.$\end{small}

where {\small $1\le i\le m$} and {\small $1\le j\le d$}.

Based on this definition, each text should have a fixed length. In this study, since two datasets with various distributions are employed, the length of each sentence is considered a fixed number. In order to meet this criterion, when there are several sentences for one image, we concatenate all the words and build a long description for that image. When the length grows to be more than the expected length, the extra words are removed and when there are fewer words, zero-padding is applied. Therefore, we will have a $70$$\times$$d$ dimension space for the representation of the sentences.

The input of the text representation model is a sequence of integer numbers. In the next step, a word embedding is used to reduce the number of semantically similar words or to remove the words with low frequency, which are non-existent in the dictionary, resulting in a new embedding space. Since the vocabulary size is very large, the reduction is helpful in increasing the network’s generalizability. The new embeddings are then fed into an LSTM \citep{gers1999learning} to learn a probability distribution over the above-mentioned sequence in order to predict the next word. The output of the LSTM is not used for word-level labeling. Instead, for the representation of the whole text, only the last hidden state is utilized. Therefore, for the input sentence T, its text descriptor denoted by $f_{txt}$ and using the function $F_{txt}(.)$, is computed as $f_{txt} = F_{txt}(T)$. The final word feature vector ($v_{txt}$) is defined by $f_{txt}(.)$.

\begin{table*}[t]
	\begin{small}\centering
	\begin{tabular*}{1\textwidth}{@{\extracolsep{\fill} } l | c c c c | c c c c | r }
	Task & \multicolumn{4}{c|}{Sentence Retrieval} & \multicolumn{4}{c|}{Image Retrieval}\\ \cline{1-9}
	Method   		& R@1	& R@5	& R@10	& Med \textit{r}& R@1	& R@5   & R@10  & Med \textit{r} & \multirow{16}{*}{\rotatebox[origin=c]{90}{1K test images}} \\ \hline
	Random Ranking 								& 0.1	& 0.6 	& 1.1 	& 631 	& 0.1	& 0.5 	& 1.0 	& 500\\
	STV \citeyearpar{kiros2015skip} 			& 33.8	& 67.7 	& 82.1 	& 3 	& 25.9 	& 60.0  & 74.6 	& 4	  \\
	DVSA \citeyearpar{karpathy2015deep} 		& 38.4 	& 69.9 	& 80.5 	& 1 	& 27.4 	& 60.2 	& 74.8 	& 3   \\
	GMM-FV \citeyearpar{klein2015associating} 	& 39.0	& 67.0 	& 80.3 	& 3 	& 24.2 	& 59.3  & 76.0 	& 4	  \\
	MM-ENS \citeyearpar{klein2015associating} 	& 39.4  & 67.9 	& 80.9  & 2 	& 25.1 	& 59.8 	& 76.6 	& 4   \\
	m-RNN \citeyearpar{mao2014deep} 			& 41.0 	& 73.0  & 83.5 	& 2 	& 29.0  & 42.2  & 77.0  & 3   \\
	m-CNN \citeyearpar{ma2015multimodal}		& 42.8  & 73.1 	& 84.1  & 2 	& 32.6  & 68.6  & 82.8  & 3   \\
	HM-LSTM \citeyearpar{niu2017hierarchical} 	& 43.9  & - 	& 87.8  & 2 	& 36.1  & - 	& 86.7  & 3   \\
	SPE \citeyearpar{wang2016learning} 			& 50.1  & 79.7  & 89.2  & - 	& 39.6  & 75.2  & 86.9 	& -   \\
	VQA-A \citeyearpar{lin2016leveraging} 		& 50.5  & 80.1  & 89.7  & - 	& 37.0  & 70.9  & 82.9  & -   \\
	2WayNet \citeyearpar{eisenschtat2017linking}& 55.8  & 75.2  & - 	& - 	& 39.7  & 63.3 	& - 	& -   \\
	sm-LSTM \citeyearpar{huang2017instance} 	& 53.2  & 83.1  & 91.5  & 1 	& 40.7  & 75.8  & 87.4  & 2   \\
	RRF-Net \citeyearpar{liu2017learning} 		& 56.4  & 85.3  & 91.5  & - 	& 43.9  & 78.1  & 88.6  & -   \\
	VSE++ \citeyearpar{faghri2017vse++}			& 64.6 	& 90.0 	& 95.7 	& 1 	& 52.0 	& 84.3 	& 92.0 	& 1   \\
	SCAN \citeyearpar{lee2018stacked}  			& 72.7 	& 94.8 	& 98.4 	& - 	& 58.8 	& 88.4 	& 94.8 	& -   \\ \cline{1-9}
	Ours								& 47.5 	& 81.0 & 91.0 & 2	    & 48.4 	& 84.3  & 91.5   & 2   \\ \hline
	GMM-FV \citeyearpar{klein2015associating} 	& 17.3	& 39.0 	& 50.2 	& 10 	& 10.8	& 28.3 	& 40.1 	& 17  & \multirow{6}{*}{\rotatebox[origin=c]{90}{5K test images}} \\
	DVSA \citeyearpar{karpathy2015deep} 		& 16.5	& 39.2 	& 52.0 	& 9 	& 10.7 	& 29.6  & 42.2 	& 14  \\
	VQA-A \citeyearpar{lin2016leveraging}  		& 23.5 	& 50.7 	& 63.6 	& - 	& 16.7 	& 40.5 	& 53.8 	& -   \\
	VSE++ \citeyearpar{faghri2017vse++}			& 41.3 	& 71.1 	& 81.2 	& 2 	& 30.3 	& 59.4 	& 72.4 	& 4   \\
	SCAN \citeyearpar{lee2018stacked}  			& 50.4 	& 82.2 	& 90.0 	& - 	& 38.6 	& 69.3 	& 80.4 	& -   \\ \cline{1-9}
	Ours						        & 23.8 	& 53.7 	& 67.3 	& 4 	& 25.6 	& 55.1 	& 68.4 	& 3   \\
	\hline
	\end{tabular*}\end{small}
	\caption{\label{table:table1}Image and sentence retrieval results on MS-COCO. ``Sentence Retrieval" denotes using an image as query to search for the relevant sentences, and ``Image Retrieval" denotes using a sentence to ﬁnd the relevant image. R@K is Recall@K (high is good). Med \textit{r} is the median rank (low is good).}
\end{table*}

\subsection{Alignment Objective}
Having an aligned collection of image-text pairs, the goal is to learn the image-text similarity score denoted by $S(T,I)$ which is defined as follows:
\begin{center}
  {\small $S(T, I) = -E(v_{txt}, v_{img})$}
\end{center}

where $v_{img}$ and $v_{txt}$ are the same-size image and text representations which have been projected onto a partial order visual-semantic embedding space. The penalty paid for every true pair of points that disagree is {\small $E(x,y) = || max(0, y-x) ||^{2}$}.

To compute the training loss, the image and text output vectors $(v_{img}, v_{txt})$ have been forced to be in the $\mathbb{R}^{+}$. By merging the image and text embedded models as illustrated in Fig.~\ref{fig:fig2}, we achieve the desired visual-semantic model. To learn an order encoding function, we considered a hinge-based triplet loss function which encourages positive examples to have zero penalty, and negative examples to have penalty greater than a margin:

\begin{small}\begin{center}
	$\sum_{(T,I)}(\sum_{T^{'}} (max\{0,\alpha - S(T,I) + S(T^{'},I)\} - \sigma^{2}(T^{'}))$\\
	$+ \sum_{I^{'}} (max\{0, \alpha - S(T,I) + S(T,I^{'})\}))  - \sigma^{2}(I^{'}) )$
\end{center}\end{small}

where $S(T, I)$, the similarity score function, is as described above while $T^{'}$ and $I^{'}$ are inferred from the ground truth by matching contrastive images with each caption and the reverse. $\sigma^{2}(x)$ is discrete variance written as {\small $\sum_{n}\dfrac{x - \mu_{c}}{|n|}$}. For computational efficiency, rather than summing over all the negative samples, we assumed only the negatives in a mini-batch.

\section{Experiments}

\subsection{Implementation}
The proposed method has been implemented with the TensorFlow \citep{abadi2016tensorflow}, and Python ran on a machine with GeForce GTX 1080 Ti. For initialization, the GloVe \citep{pennington2014glove} word embeddings, trained on Twitter with 1.2 million vocabulary size, 27 billion tokens and 2 billion tweets, are employed. The training phase starts with an Adam optimizer with learning rate of 0.1 and a batch size of 16 and continues as long as the amount of loss does not change. When it happens, the learning rate is divided by 2. This continues until the learning rate becomes $10^{-7}$. Then, the batch size is doubled and the learning rate is reset to 0.1. We repeat this process to optimize the model. During the training, a grid search over all the hyper-parameters is carried out in order to conduct a model selection. For efficiency, the training is performed in batches which allows us to do real-time data augmentation on images in CPU in parallel with training the model in GPU.

\subsection{Evaluation}
Given a sentence (image), all the images (captions) of the test set are retrieved and listed based on their penalty in an increasing order. Then we report the results using \textit{Recall} and \textit{Median Rank}. \textit{Recall} is a metric for assessing how well a system retrieves information to a query. It is computed by dividing the number of relevant retrieved results by the total number of instances. In $R@K$, the top $K$ results are treated as the output and the \textit{Recall} is computed accordingly. \textit{Med} $r$ is the middle number in a sorted sequence of the retrieved instances.

To address this issue, other metrics can be taken into account since the existing measures can be intrinsically problematic \citep{bernardi2016automatic}. For instance, the retrieval of the exact image (text) is not guaranteed. In these cases, since the exact matches have not been retrieved, its score is considered although similar ones have been matched.

\subsection{Data Collection and Results}
Several datasets have been published for image-sentence retrieval task \citep{rashtchian2010collecting, ordonez2011im2text, young2014image, hu2017twitter100k, farhadi2010every}. We collect a dataset, as a proof of concept, for evaluating and analyzing our method to better showcase its ability to generalize as well as for demonstrating the extensibility of this type of solution to conversational texts and unusual images.
Moreover, we used MS-COCO \citep{lin2014microsoft} to train and test the proposed model. This dataset contains 123,287 images and 616,767 descriptions \citep{lin2014microsoft}. Each image contains 5 textual descriptions on average which collected by crowdsourcing on AMT. The average caption length is 8.7 words after rare word removal. We follow the protocol in \citep{karpathy2014deep} and use 5000 images for both validation and testing, and also report results on a subset of 1000 testing images in Table~\ref{table:table1}.

We collected 13751 tweets with 14415 images by a crawler based on the Twitter API. To make sure that the collection is diverse, we first created a list of seed users. Then, the followers of the seed accounts were added to the list. Next, the latest tweets of the users in our list were extracted and saved in the dataset. To make the data appropriate for our task, we removed retweets, the tweets with no images, non-English tweets and the ones that had less than three words. This led the dataset to have a relatively long description for each image and at least one image for every tweet. At the final step, the dataset was examined by two professionals and unrelated content was removed by them. Fig.~\ref{fig:fig1} shows samples of the extracted dataset. This collection is different from currently existing ones due to varied domains, informal texts and high level correlation between text and image. For instance, the tweets may contain abbreviations, initialisms, hashtags or URLs. On collected tweets, our model improves sentence retrieval by 14.3\% relatively and image retrieval by 16.4\% relatively based on $R@1$. The dataset has been available.\footnote{Dataset, source codes and model will be publicly available after publishing the paper.}

\section{Discussion}
We propose a multi-modal image-text matching model using a convolutional neural network and a long short-term memory along with fully-connected layers. They are employed to map image and text inputs into a shared feature space, where their representations can be compared, to find the closest pairs. Additionally, a new dataset of images and tweets extracted from Twitter is introduced, with the aim of having a characteristically different collection from the benchmarks. Whereas the descriptions in the benchmarks are well-organized, our dataset has not been standardized and the image-text pairs can contain high semantic correlations. Also, because of a varied number of domains existent in the extracted dataset, the task of image-text matching becomes even more challenging. Therefore, it can be used to carry out new research and assess the robustness of the proposed frameworks. Our experiments on MS-COCO yield improved results over some previously proposed architectures.

\ifaclfinal
\section*{Acknowledgments}

The research was partially supported by OP RDE project No. CZ.02.2.69/0.0/0.0/16\_027/0008495, International Mobility of Researchers at Charles University.
This project was in part supported by a grant from Institute for Research in Fundamental Sciences (IPM). \\
We gratefully acknowledge the support of NVIDIA Corporation with the donation of the GPU used for this research.\\
\fi

\bibliography{DeepMulRetPaper}
\bibliographystyle{acl_natbib}

\end{document}